# Design of a prototype device to calibrate the Large Size Telescope camera of the Cherenkov Telescope Array

**M. Iori[1]**

*1 Department of Physics, Sapienza University of Rome*
*P.zza A. Moro 5 00185 Rome Italy*
*E-mail: maurizio.iori@roma1.infn.it*

**P. Majumdar [2], F. De Persio [4], A. Chatterjee [2], F. Ferrarotto[4], B. K. Nagesh [3], L. Saha [3], B. B. Singh [3]**

*2  Saha Institute of Nuclear Physics, 1/AF Bidhannagar, Sector-I Kolkata-700064, India*
*3 Tata Institute of Fundamental Research, Mumbai, India*
*4 INFN Sezione di Roma1, Piazzale Aldo Moro 2, 00185, Rome, Italy*

**for the CTA Consortium**[*]

The Cherenkov Telescope Array is a project that aims to exploring the highest energy region of electromagnetic spectrum. Two arrays, one for each hemisphere, will cover the full sky in a range from few tens of GeV to hundreds of TeV improving the sensitivity and angular resolution of the present operating arrays. A prototype of the Large Size Telescope (LST) for the study of gamma ray astronomy above some tens of GeV will be installed at the Canary Island of La Palma in 2016. The LST camera, made by an array of photomultipliers (PMTs), requires an accurate and systematic calibration over a wide dynamic range. In this contribution, we present an optical calibration system made by a 355 nm wavelength laser with 400 ps pulse width, 1 µJ output energy, up to 4k Hz repetition rate and a set of neutral density filters to obtain a wide range of photon intensities, up to 1000 photoelectrons/PMT, to be sent to the camera plane 28 m away. The number of photons after the diffuser of the calibration box, located in the center of the reflective plane, is monitored by a photodiode. The stability of the laser and the ambient parameters inside this calibration box are checked by a multi-task processor and a trigger signal is sent to the camera data acquisition system. The box frame is designed with special attention to obtain a robust device with stable optical and mechanical features.

---

[1]Speaker
[*]full Consortium author list at https://www.cta-observatory.org/

1. Introduction

The Cherenkov Telescope Array, CTA, is an observatory for Very-High-Energy, VHE, gamma-ray astronomy which will provide observers with data on astrophysical objects over a very wide range of gamma ray energies. CTA will exploit the Imaging Atmospheric Cherenkov Technique, IACT, to measure the energy, direction and arrival time of gamma-ray photons arriving at the Earth from astrophysical sources. CTA will consist of an array of telescopes at a given site, with UV-optical reflecting mirrors focusing flashes of Cherenkov light produced by atmospheric particle cascades on their camera. CTA will cover an energy range extending over five orders of magnitude in energy and desired the use of different telescope sizes to achieve the required sensitivity at acceptable cost. Three telescope types are planned for the baseline array to achieve high sensitivity in a wide energy range (100 MeV-100 TeV): a compact group of four Large Sized Telescopes, LSTs, Medium Sized Telescopes, MSTs, distributed across a 1 km$^2$ area, and Small Sized Telescopes, SSTs, distributed across several square kilometers. Each LST telescope is planned to have a parabolic reactive surface sustained by a tubular structure to collect and focus the Cherenkov radiation into the Camera, where a Photo-Multipliers Tube (PMT) matrix converts the light in electrical signals that can be processed by dedicated electronics. The mirror has a diameter of 23 m and a focal length of 28 m. LSTs are arranged at the center of arrays to lower the energy threshold and to improve the sensitivity of CTA below 200 GeV. The low energy threshold provided by the LSTs is critical for CTA studies of galactic transient, high redshift Active Galactic Nuclei and Gamma Ray Bursts.

A periodic calibration is necessary before starting data acquisition to achieve precise measurements. For this task we propose a Calibration Box, named in the text CaliBox, placed at 24 m from the Camera, in the center of the parabolic mirror structure. CaliBox illuminates the Camera with a UV light source of known intensity and determines the ratio between the number of digital counts recorded by the Readout System and the photo-electrons (phe's). In order to calibrate the camera, the absolute gain of each electronic channel has to be accurately evaluated and the light detected by the pixel corrected.

The laser light source produces a 400 ps pulse of UV light, arbitrarily triggered. The intensity of the light can be modulated by absorptive neutral density filter. Finally, the light beam is widespread to the whole Camera area through a diffuser inside the CaliBox. All the components and sensors of the calibration box will be controlled using a single board computer ODROID-C1. The UV laser rate must be adjustable between 1 Hz and 2000 Hz. Pulse intensity fluctuations must be less than 5% for all the 30 minutes operation time expected to perform the Camera calibration. Filters combinations must ensure a dynamic range from 1 to 80000 photo-electrons to the Camera PMT. The diffuser must ensure the uniformity of light within 3% at the Camera plane. These parameters have been already tested with the calibration system build for Magic Telescope.

The controller must able to run the Open Platform communication unified architecture, OPC-UA, to communicate to a server. CaliBox need to be dust and water proof and have an optimal thermal design in order to dissipate all the heat generated by its internal components. In the section 2 we give description of the electronics, optics and the frame housing the hardware.

## 2. Scheme of the CaliBox

### 2.1 *Electronics*

Figure 1 shows the scheme of the electronics. The main core of CaliBox is the single board computer ODROID-C1 produced by HardKernel company. It is from the same family of the well know Raspberry Pi. The board is equipped with a quad core Amlogic S805 processor based on ARMv5 cores clocked at 1.5 Ghz, 1 Gb of LP (low power) DDR3 RAM, 40 generic purpose input/output pin (GPIOs), 4 USB ports, 1 Gigabit Ethernet port and support both SD and eMMC card. GPIOs support SPI and I2C communication standards and power pins can give 3.3V and 5V to enhance compatibility with peripherals, devices, sensors or Arduino based shields. The board was chosen for three main features: - it has a low power consumption with an expected consumption of 2.5 W, - the board can run a full Ubuntu Linux, - the S805 chip integrates two 1.8 V SAR ADCs ( Serial Approximation Register Analog To Digital Converter) at 10 bit capable of 8k samples per second (ksps). The main tasks of the ODROID-C1 Board are: a) control the filter Thorlabs FW102C wheels in order to provide different flux of photons, b) to send the start and stop signal to the laser controller, c) to read the temperature and humidity using a HardKernel weather-station sensors in the CaliBox, d) to acquire data from the photodiode placed at the edge of the diffuser. The UV Teemphotonics STV-01E-1x0 laser is controlled using the RS232 standard protocol through and USB-RS232 adapter. The Centronic 5.8 mm$^2$ quartz window photodiode optimized in the range 196 nm to 400 nm is connected to the diffuser and read with a frontend by ADC to monitor the photons flux at exit of diffuser. With a full Ubuntu Linux operating system on the board, the C1 is capable to run the OPC-UA server with no limitations or restrictions.

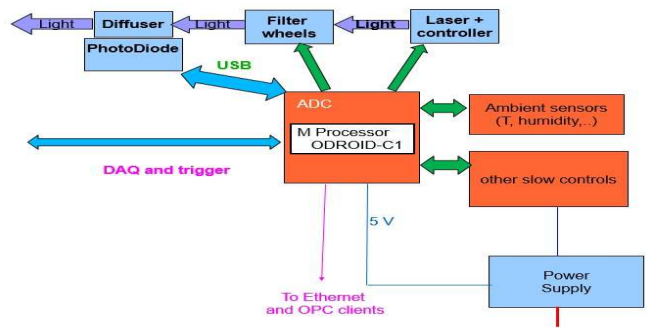

Figure 1: Scheme of the tasks performed by the ODROID-C1 board.

## 2.2 *Optics*

As already mentioned in [1], according to the Technical Design Report (TDR) [2] requirements, the laser should provide a light density at the camera plane of 10 phe to 1000 phe. A stability of the flash amplitude is required on a minute scale. To meet these requirements we choose a Teem-photonics STV-01E-1x0 UV laser with a wavelength of 355 nm, programmable repletion rate up to 4 KHz, pulse width of 0.4 ns and a single output energy beam greater than 1 µJ and a peak power of 2.5 kW. The maximum power consumption is 40 W (15 W from the laser head) at full repetition rate. Considering also the laser controller, the maximum power consumption is below 45 W. The heat dissipation system is delegated to an optimized bracket that connects the laser to the aluminum box base. To reach a dynamic range of photons on the camera plane, the laser beam need to be attenuated by a combination of filters. For this purpose we use the Thorlabs Neutral Density, ND, filters with the Optical Density, OD, in the interval 0.2 and 4.0 and working in a range of 350 nm to 700 nm. Calibox has two 6-position motorized wheels for 2.5 cm diameters filters for a total of 36 filters combinations, connected and driven by the ODROID-C1 board with an USB connection. We have performed several tests to measure the variation of transmission as function of OD respect to the nominal value provided by the manufacture as shown in Figure 2. The variations found in the measurements of the performance of individual filters with respect to the manufacturer characterization is taken into account to evaluate the dynamic of the photo-electrons request for each pixel that ranges between few to 1000 phe. The current setup does not allow to measure the transmission coefficient at wavelength of 355 nm for OD bigger than 2.0 due to limited sensitivity of photodiode. Another component of the optics is the diffuser that spreads the beam laser uniformly within 2% over the camera plane. To perform this task we have developed a diffuser made of Teflon with a cavity of 10 mm diameter and an exit of 2 mm diameter and an open angle of about $5^0$. With this geometry 1/100 of the incident light will go through the exit hole and it illuminates a circle area with a diameter of 5.6 m at 24 m of distance, covering the whole area of the LST Camera. A second light cone, with same size of the outlet cone is used to evaluate the photons sent to the LST camera using a photodiode. The laser yields $1.7 \times 10^{12}$ photons per pulse that correspond without any filters to $6.9 \times 10^4$ photons/cm$^2$ at the Camera Plane. With maximum filters OD the number of photons drops to $5.3 \times 10^3$ photons at the diffuser entrance and $2.1 \times 10^{-4}$ photons/cm$^2$ on the camera plane.

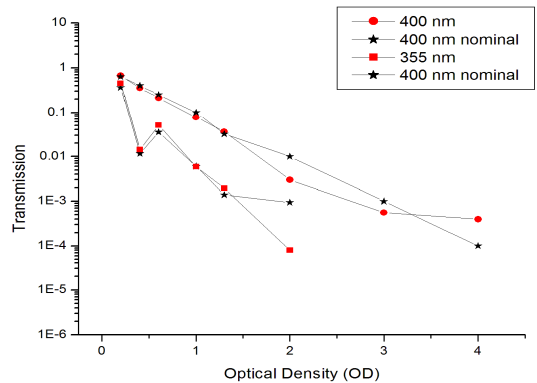

Figure 2: Measurements of transmission (dot and box) versus different ODs at two wavelengths, 400 nm and 355 nm, compared to the nominal values (stars) provided by the Thorlabs [3]. Errors are not visible in this scale.

### 2.3 *The frame of CaliBox*

The frame of the CaliBox, as shown in Figure 3, consists of a 41 cm x 31 cm x 0.5 cm aluminum plate. The aluminum plate works as a thermal radiator to dissipate the heat generated by all the active components. The UV laser and wheels are placed together inside an aluminum box connected to the diffuser box in order to reduce the thermal excursion and the dust on the filters and guaranty stable optical performance. In order to reduce the humidity and the dust as much as possible inside the box, a vacuum valve is connected to the Laser and wheel box in order to create a depletion of air inside the boxes of the optics path. A study of the thermal dissipation of the laser has been also performed. Outside the optics boxes, there is the 100 W power supply, a Tracopower TXL-100 12S and the ODROID-C1 computer board. All the cables, USB and power cables and the connections ports, from the optics boxes and the power supply and controller board placed outside have IP67 standard connectors. The optics boxes, power supply and ODROID-C1 will be covered by a plastic box, in order to protect the hardware from the atmospheric agents. The estimated weight of the CaliBox frame is about 13 kg and including the devices its weight is about 16 kg.

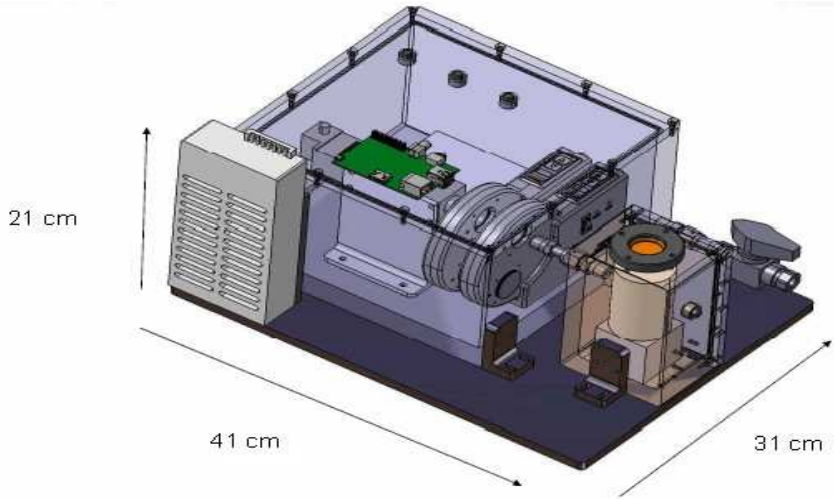

Figure 3: Scheme of the CaliBox described in sec.2.3. On left side is visible a box containing the Laser and two filter wheels connected to the box with the diffuser. Both boxes are closed by a vacuum valve to remove air from inside. The ODROID-C1 and the power supply are placed on top and on side of the box containing the Laser. The cables and the connectors are not drawn.

## 3. *Conclusion*

The CTA is a proposed new generation of gamma ray observatory with high sensitivity. To achieve precise measurements of astronomical phenomena a calibration procedure is necessary. This paper describes a prototype of a calibration device for the camera of the Large Size Telescopes proposed for CTA. It uses a laser with a flux of about $1.7 \times 10^{12}$ photons and an optical system that provides a uniform density of photons on the camera plane. We aim for an accuracy of 2% in uniformity of light intensity. This prototype will be ready and tested in the fall this year.

ACKNOWLEDGMENTS

We gratefully acknowledge support from the agencies and organizations listed under Funding Agencies at this website: http://www.cta-observatory.org/. We would also like to thank Prof. M. Capizzi, D. Tedeschi of Solid State Physics Laboratory, University La Sapienza, Dr. L. Recchia and Dr. A. Zullo of INFN-RM1 Electronics Laboratory (LABE) and INFN-RM1-Engineering Design Office.